\RequirePackage{amsmath}
\documentclass[runningheads,a4paper]{llncs}
\usepackage{amssymb}
\usepackage{graphicx}
\usepackage{url}
\newcommand{\keywords}[1]{\par\addvspace\baselineskip
\noindent\keywordname\enspace\ignorespaces#1}
\usepackage[T1]{fontenc}
\usepackage[utf8]{inputenc}
\graphicspath{{}}
\usepackage{algpseudocode}
\usepackage{array}
\usepackage{ifthen}
\usepackage{algorithm}
\usepackage[caption=false]{subfig}
\usepackage{multirow}               
\usepackage{amsmath}
\usepackage{color}
\usepackage{cite}
\usepackage[margin=1in]{geometry}
\usepackage{rotating}
\usepackage{framed}
\usepackage[table,xcdraw]{xcolor}
\def\quad{\hskip1em\relax}
\usepackage[nonumberlist,acronym,toc,section]{glossaries}
\usepackage{mwe}    
\usepackage{subfig}
\providecommand{\keywords}[1]{\textbf{\textit{}} #1}
\begin{document}

\mainmatter  

\title{On the Ability to Reconstruct Ancestral Genomes \\
from \textit{Mycobacterium} Genus}

\titlerunning{On the ability to reconstruct ancestral genomes from \textit{Mycobacterium} genus}

%
%
\author{C. Guyeux\inst{1} \and B. Al-Nuaimi\inst{1,2} \and B. AlKindy\inst{3} \and J.-F.~Couchot\inst{1} \and M. Salomon\inst{1}}
\institute{$^1$~FEMTO-ST Institute, UMR 6174 CNRS, DISC Computer Science Department, \\ Univ. Bourgogne Franche-Comt\'{e} (UBFC), France\\
$^2$Department of Computer Science, University of Diyala, Iraq\\
$^3$~Department of Computer Science, University of Mustansiriyah, Baghdad, Iraq\\
\email{christophe.guyeux@univ-fcomte.fr}}

\toctitle{Lecture Notes in Computer Science}
\tocauthor{Authors' Instructions}
\maketitle

\begin{abstract}
Technical signs of progress during the last decades has led to a situation in which the accumulation of genome sequence data is increasingly fast and cheap. The huge amount of molecular data available nowadays can help addressing new and essential questions in Evolution. However, reconstructing evolution of DNA sequences requires models, algorithms, statistical and computational methods of ever increasing complexity. Since most dramatic genomic changes are caused by genome rearrangements (gene duplications, gain/loss events), it becomes crucial to understand their mechanisms and reconstruct ancestors of the given genomes. This problem is known to be NP-complete even in the ``simplest'' case of three genomes. Heuristic algorithms are usually executed to provide approximations of the exact solution. 
We state that, even if the ancestral reconstruction problem is NP-hard in theory, its exact resolution is feasible in various situations, encompassing organelles and some bacteria. Such accurate reconstruction, which identifies too some highly homoplasic mutations whose ancestral status is undecidable, will be initiated in this work-in-progress, to reconstruct ancestral genomes of two \textit{Mycobacterium} pathogenetic bacterias. By mixing automatic reconstruction of obvious situations with human interventions on signaled problematic cases, we will indicate that it should be possible to achieve a concrete, complete, and really accurate reconstruction of lineages of the  \textit{Mycobacterium tuberculosis} complex. Thus, it is possible to investigate how these genomes have evolved from their last common ancestors.
\end{abstract}

\section{Introduction}

\textit{Mycobacterium tuberculosis} is presently still one of the principal causes of death worldwide. Approximately one-third of the world population is infected by the \textit{Mycobacterium tuberculosis complex} (MTBC), with about 9 million event cases annually, leading to estimated a million deaths each year. Due to their different host tropism and phenotypes, members of MTB complex display various pathogenicities ranging from particularly human 
(\textit{M. tuberculosis}, \textit{M. africanum}, and \textit{M. canetti}) or rodent pathogens (\textit{M. microti}) to~\textit{Mycobacteria} with a broad host spectrum (\textit{M. bovis})~\cite{smith2006bottlenecks, shamputa2015introduction, brosch2002new}. 
\textit{Mycobacterium tuberculosis} has been in the human population around for thousands of years, as fragments of the spinal column of Egyptian mummies from 2300 BCE show definite pathological signs of tubercular decay. It has been recognized as the leading cause of mortality by 1650, while using a new staining technique, Robert Koch identified the bacterium responsible for causing consumption in 1882. 

The MTB complex belongs to the slow-growing sublineage of \textit{Mycobacteria}. Based on  topographical characteristics, MTBC can be categorized into six clusters, including species such as~\textit{M. tuberculosis}, \textit{M. africanum},  \textit{M. bovis},  \textit{M. microti}, and \textit{M. canettii}. Members in MTBC share $99.95\%$ of their genomic sequences and a rigorously clonal population structure~\cite{gutacker2002genome}. Compared to more ancient species (\textit{e.g.}, \textit{M.~marinum}), MTBC has shorter but more virulent chromosomes~\cite{mostowy2002genomic,yamada2007mycobacterium}. Considering that they all are derived from a common ancestor, it is interesting that some are human or rodent pathogens, whereas others have a wide host spectrum~\cite{fabre2010molecular}.
The genome of \textit{M. tuberculosis} was studied using the strain 
\textit{M.~tuberculosis H37Rv}. It has a circular chromosome of about 4,200,000 nucleotides long, while containing about 4,000 genes~\cite{fleischmann2002whole}. The different species of the \textit {Mycobacterium tuberculosis} complex show a $95-100\%$ DNA relatedness based on studies of DNA homology, and the sequences of the 16S rRNA gene are the same for all the species. 

MTBC genomes have been modified during the evolution by mutation, insertion-deletion of nucleotides, by large-scale changes (inversion, duplication or deletion of large DNA strands), or by other modifications specific to repetition (insertion sequences, etc.). 
Being able predict both its past or its future evolution 
may have multiple applications: 
to reconstruct the past history and the ancestors of bacteria, or to better understand their mechanism of virulence and resistance acquisition.
The relatively short timescale (tuberculosis disease is relatively recent, as its most recent common ancestor evolved $\approx$ 40,000 years ago~\cite{wirth2008origin}), the relatively reasonable sizes of considered genomes, the relative rarity of recombination events, and the recent possibility to have access to old and present bacterial DNA sequences, may lead to the possibility to model the evolution of these genomes, in order to reconstruct and to understand their ancient history and to predict their future evolution. 

To do so, new algorithms of detection and of evolution regarding genomic modifications must be written. People working on this problematic mainly focus on predicting the evolution of nucleotide mutations, and by assuming specific forms for matrix mutations which seem incompatible with recent experimental measures~\cite{lang2008estimating}. These models for evolution must be designed differently, in order to better reflect the reality. Additionally, the serious impact of other modifications operating on the genomes (as insertions and deletions of nucleotides (indels), inter and intra chromosomic recombinations, or modifications specific to repetition), must be taken into account more deeply, while a concrete ancestral reconstruction of bacterial lineage must be finally achieved.

The objective of this work-in-progress is to prove that, given a set of close bacterial genomes, it is possible to reconstruct in practice their recent sequence evolution history, by mixing state-of-the-art tools with a pragmatic manual completion and cross-validation. 
We will illustrate that, in practice, it should be possible to reconstruct ancestral genomes for some lineages of the  \textit{Mycobacterium} genus, using all available complete genomes of such a lineage (for instance, 65 complete genomes of the MTB complex are currently available, and we have more than 1,000 archives of reads).

The remainder of this article is organized as follows. 
In Section~\ref{Background}, we start by giving reviews of computational approaches and tools for analyzing the evolution of DNA sequences. 
We propose in the next section a set of methodological principles that can be used for ancestral genome reconstructions, and how to apply them on \textit{M. canettii} and \textit{M. tuberculosis} data. Obtained results and further perspectives are discussed in Section~\ref{results}. This research work ends with a conclusion section in which the article is summarized and intended future work is outlined.

 \section{Scientific background}\label{Background}

\begin{figure*}
\centering
\includegraphics[width=10 cm]{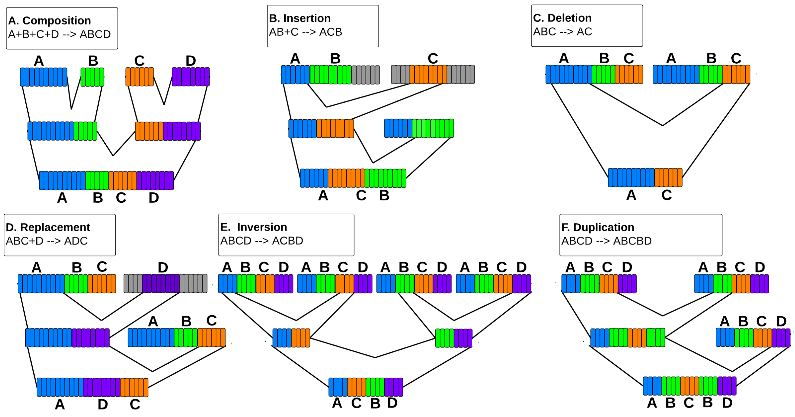}
\caption{Various genome rearrangement events.} 
\label{fig:genome rearrangement}
\end{figure*}

\subsection{On genomic evolution}
It is well-known that DNA sequences change over time due to local mutations, which are either single nucleotide polymorphism (SNP) or insertion-deletion (indel) of one nucleotide.
Mutations that affect the organization of genes are called genome rearrangements, 
which include inversions, transpositions, and chromosome fusions and fissions. 
Example of such large scale modifications are illustrated in Figure~\ref{fig:genome rearrangement}.
During evolution, such large-scale mutations rearranging the genome have occurred, and both gene order and content have been modified accordingly, which may
represent a meaningful role in speciation~\cite{fertin2009combinatorics}.

One important problem in molecular evolution, which is targeted by this study, is that of reconstructing ancestral genomic sequences. 
In this problem, an evolutionary tree of organisms is provided, together with genomic sequences for the leaf species. 
The aim is to infer the genomic sequences of the ancestral nodes in the tree, that is, those of the organisms that no longer exist. 
Various methods have been developed to infer such ancestral sequences and they already have been used in various biological studies. 

More precisely, evolution of biomolecules over time have mainly been computationally studied in two directions, namely through ancestral genome reconstruction problem and through the evolution of pan and core genomes over time. A short overview of these topics is provided below.

\subsection{Ancestral genome reconstruction}

Ancestral reconstruction may focus at sequence level or at gene order level, the former being quite resolved~\cite{ma2008dupcar,gagnon2012flexible,jones2012anges,ma2006reconstructing,10.1109/TCBB.2014.2309602,blanchette2008computational,rascol2007ancestral,larget2005bayesian,hannenhalli1995genome}, at least if we do not consider indels and mutation neighborhood, while the latter is more difficult in general, due to its combinatorial complexity. More precisely, given an alignment of DNA sequences and a tree, ancestral nucleotides of extant species can be obtained by modeling the evolution of a trait through time as a stochastic process (Markov chain). Using it as the basis for statistical inference, both maximum likelihood or Bayesian inference approaches can be applied to estimate ancestral configuration. 

Well known software like RAxML~\cite{stamatakis2014raxml}, BEAST2~\cite{bouckaert2014beast}, or PAML~\cite{yang2000phylogenetic} can be used for such reconstructions. However, most of the time, like in the R package~\cite{paradis2004ape}, indels are not considered in such ancestral state reconstruction, even if researches have recently been realized via the so-called ``Poisson Indel Process''~\cite{bouchard2013evolutionary}. Such process is a significant improvement, if we compare it with the parsimony approach that can be found in PHAST software, or with the Thorne-Kishino-Felsenstein model of indel evolution. Large scale modifications, for its part, is most of the time regarded in a combinatorial framework by modeling genomes as permutations of genes or homologous regions. Indeed, this genome rearrangement problem~\cite{watterson1982chromosome} is usually formulated as follows: ``given two genomes (permutations) and a set of allowable operations (like inversion, deletion, or transposition), what is the shortest sequence of operations that will transform one genome into the other?''. However, even in the case of three genomes, such a problem is NP-hard~\cite{even1981minimum}, although it has received much attention in mathematics and computer science~\cite{fertin2009combinatorics}.

An important remark, motivating our proposal, is that the NP-hard character of this problem only appears if we consider a very large number of operations in very large sequences. On our side, we will consider quite small sequences and a relatively small number of large scale recombinations. So we face tractable problems in various real situations, on which simple and pragmatic approaches may work.

\subsection{Core and pan genome extraction}
An early study about finding the common genes in chloroplasts has been realized by Stoebe et al. in 1998~\cite{stoebe1998distribution}. They established the distribution of 190 identified genes and 66 hypothetical protein-coding genes (ysf) in all nine photosynthetic algal plastid genomes available (excluding non-photosynthetic Astasia tonga) from the last update of plastid genes nomenclature and distribution. The distribution reveals a set of approximately 50 core protein-coding genes retained in all taxa. In 2003, Grzebyk et al.~\cite{grzebyk2003mesozoic} have studied the core genes among 24 chloroplast sequences extracted from public databases, 10 of them being algae plastid genomes. They broadly clustered the 50 genes from Stoebe et al. into three major functional domains: (1) genes encoded for ATP synthesis (atp genes); (2) genes encoded for photosynthetic processes (psa and psb genes); and (3) housekeeping genes that include the plastid ribosomal proteins (rpl and rps genes). The study shows that all plastid genomes were rich in housekeeping genes with the rbcL gene involved in photosynthesis. Other examples of such core and pan studies can be found in, e.g.,~\cite{sharon2009photosystem,de2014genome,kurtz2004versatile}.

Concerning bacterias, many studies have recently achieved the extraction of core and pan genomes using NCBI annotations, which are mainly based on generic annotation tools like Glimmer, GeneMarkS, or Prodigal (see for instance~\cite{touchon2009organised,boissy2011comparative,tettelin2005genome}, the Pseudomonas aeruginosa case being resolved by us in~\cite{valot2015takes}). In most of these studies, considered genomes have been annotated with various different annotation algorithms, mixing human curated and automated coding sequence prediction tools that are not specific to the genus under consideration. This large variety of manners to detect coding sequences and their functionality leads to large variability in gene boundaries (start and stop codons) and naming process, which obviously severely biases the core and pan genomes determination.

\section{A concrete semi-automatic ancestral reconstruction}
\label{phylogenetic analysis}

\subsection{General presentation}
\label{Problem review}
By a phylogenetic study, it is possible to reconstruct the evolutionary relationship of a set of organisms in the form of a binary tree, in which the given set of organisms are descendants placed at the leaves, while internal nodes stand for extinct ancestors connected by edges. We argue that, knowing this tree, ancestral genomes can be completely reconstructed in some easy cases, by aligning extant genomes and finding homologies between them, and then inferring various scenarii of evolutionary events during history~\cite{yang2011analysis}. This ancestral reconstruction can be achieved by mixing state-of-the-art algorithms and manual investigations, if the considered genomes have not evolved so much. To illustrate the feasibility of the proposal, an example of such reconstruction is provided in this section, in the case of the MTB complex.

\begin{figure}
\centering
\includegraphics[width=5 cm]{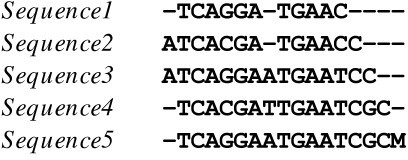}
\caption{Representation of a multiple sequence alignment.} 
\label{fig:Align}
\end{figure}

\begin{table}[!htb]
\caption{Information about some \textit{Mycobacterium} genomes}
\tiny
\begin{center}
\begin{tabular}{l|c|c|c}
\multicolumn{1}{c|}{Organism name} & Accession & Sequence length & Number of genes\\ \hline\hline
\textit{Mycobacterium tuberculosis W-148} & NZ\_CP012090.1 & 4,418,548 bp & 4,133\\
\textit{Mycobacterium tuberculosis H37Rv} & NC\_018143.2 & 4,411,709 bp & 4,132  \\
\textit{Mycobacterium africanum GM041182} & NC\_015758.1 & 4,389,314 bp & 4,089  \\  
\textit{Mycobacterium africanum strain 25}& CP010334.1 & 4,386,422 bp & 4,798 \\ 
\textit{Mycobacterium microti strain 12} & CP010333.1 & 4,370,115 bp & 4,321 \\ 
\textit{Mycobacterium canettii CIPT 140010059} & NC\_015848.1 & 4,482,059 bp & 4,137  \\
\textit{Mycobacterium canettii CIPT 140070008} & NC\_019965.1 & 4,420,197 bp & 4,103  \\ 
\textit{Mycobacterium bovis strain ATCC BAA-935} & NZ\_CP009449.1 & 4,358,088 bp & 4,095 \\ 
\textit{Mycobacterium bovis BCG str. Tokyo 172} & NZ\_CP014566.1 & 4,371,707 bp & 4,076 
\end{tabular}
\end{center}
\label{tab:Mycobacterium}
\end{table}

To illustrate this claim, the complete sequences of 65 \textit{Mycobacterium} genomes, which are available on the NCBI\footnote{ftp://ftp.ncbi.nih.gov/genomes} 
have been downloaded. Listed according to their species, 42 genomes of \textit{tuberculosis}, 15 \textit{bovis}, 2 \textit{africanum}, 5 \textit{canettii}, and 1 \textit{microti} have been recovered. Table~\ref{tab:Mycobacterium} shows information about some of these  \textit{Mycobacterium} genomes. Among this MTBC, we particularly focused on \textit{tuberculosis} and on \textit{canettii}, as there are enough of them, and because the virulent \textit{tuberculosis} species is supposed to have emerged from \textit{canettii} forty thousand years ago. 
To verify such an evolutionary hypothesis, the first task of our approach, proposed to achieve an ancestral reconstruction of close genomes, is to perform a multiple sequence alignment of the sequences. This task is described in the next section.


\subsection{Multiple sequence alignment}

the first stage of this alignment stage, is to identify a common starting point in these complete circular genomes. In order to do so, we searched for a reference sequence of 200 nucleotides from \textit{M.~tuberculosis H37Rv}, and we found it or its transconjugate in each genome using a local blast. Then, a circular rotation (together with a transconjugate operation if needed) has been performed on each complete genome, so that each sequence starts with the same 200 nucleotides, if we except SNPs. 
Once these sequences have been operated to share the same orientation and starting location, the overall alignment of each chromosome has been performed.

Alignment of large sets of sequences is a common task during biological investigations and has a wide variety of applications incorporating homology detection~\cite{wang2009procain}, finding evolutionarily relevant sites, and phylogenetics. A multiple sequence alignment, as depicted in Figure~\ref{fig:Align}, may explain many aspects about a gene: which regions are constrained, which sites undergo positive selection~\cite{kemena2009upcoming}, and potentially the structure of its gene product~\cite{warnow2013large}. Furthermore, aligning sequences can help to detect events of mutations or recombination in couples of close genomes, which is valuable for what we intend to do. To achieve such an alignment, we thus have considered the \textit{AlignSeqs} function from Decipher R package~\cite{team2013r}. Indeed, after various tests on well known alignment tools, this latter was the only one that achieved to align complete bacterial genomes with a good accuracy.

\begin{figure}[!htb]
\centering
\includegraphics[width=13 cm]{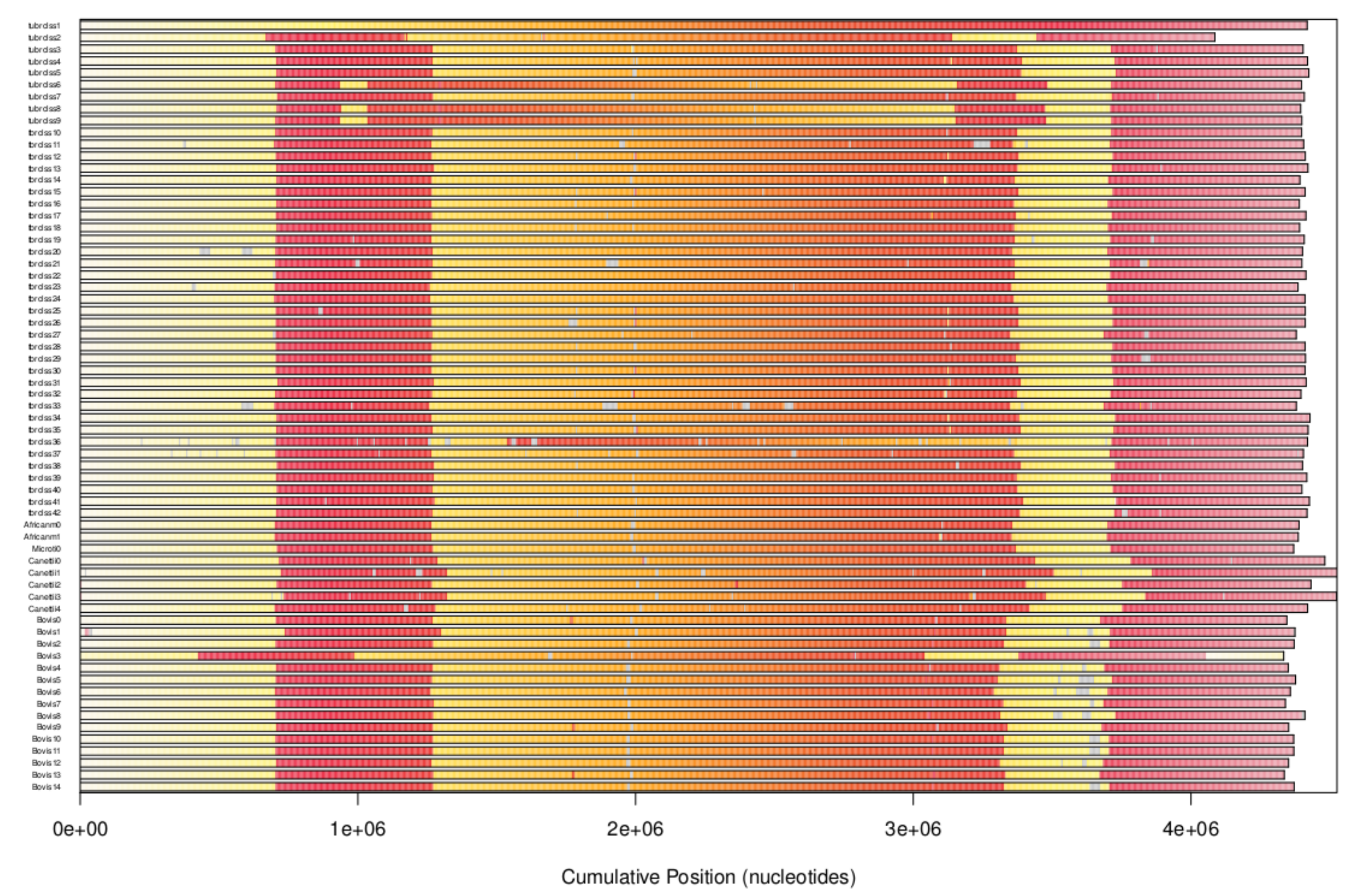}
\caption{A synteny representation of all available \textit{Mycobacterium} strains} 
\label{fig:Mycobacterium_1}
\end{figure}

This \textit{AlignSeqs} function takes as input two aligned sets of DNA sequences and returns a merged alignment. It can be used to achieve multiple sequence alignment in a progressive or iterative manner on sequences of the same kind. Indeed, multiple alignments are accomplished by aligning two sequences, merging with another sequence, combining with another set of sequences, and so on until all the sequences are aligned~\cite{gentleman5840,wright2014art}. 
We thus obtained a first representation of synteny of the whole 65 \textit{Mycobacterium} genomes, which is depicted in Figure~\ref{fig:Mycobacterium_1}. 
It can be observed that these 65 genomes have a high sequence similarity with low recombination events.

\subsection{Phylogenetic study}

This first representation of synteny blocks, obtained thanks to the multiple sequence alignment of the whole \textit{Mycobacterium} genus, has allowed us to detect the location of a few large scale inversions. We thus have been able to manually invert again these inversions, so that the multiple alignment became quite perfect, if we except small indels and SNPs. It is then possible to use all the 65 complete genomes in the next stage, namely the phylogenetic study.

Indeed, the evolutionary history of our population of genomes can be represented as a 
phylogenetic tree using the multiple sequence alignment combined with manual 
local inversions previously obtained.  
Various methods are well established in the literature to investigate the best phylogenetic tree for a given set of aligned sequences. 
Well-known techniques for phylogenetic analysis include parsimony methods, maximum likelihood, distance-based methods, and even artificial intelligence based ones~\cite{agcs+15:oip, alkindy2015binary}.
On our side, we decided to consider the use of RAxML as a default phylogenetic tree reconstruction toolkit, a well known and reputed software based on maximum likelihood~\cite{alkindy2014hybrid,stamatakis2014raxml}.

\begin{figure}[!htb]
\begin{minipage}{.5\linewidth}
\centering
\subfloat[]{\label{fig:M.canettii_tree}\includegraphics[width=0.8\textwidth]{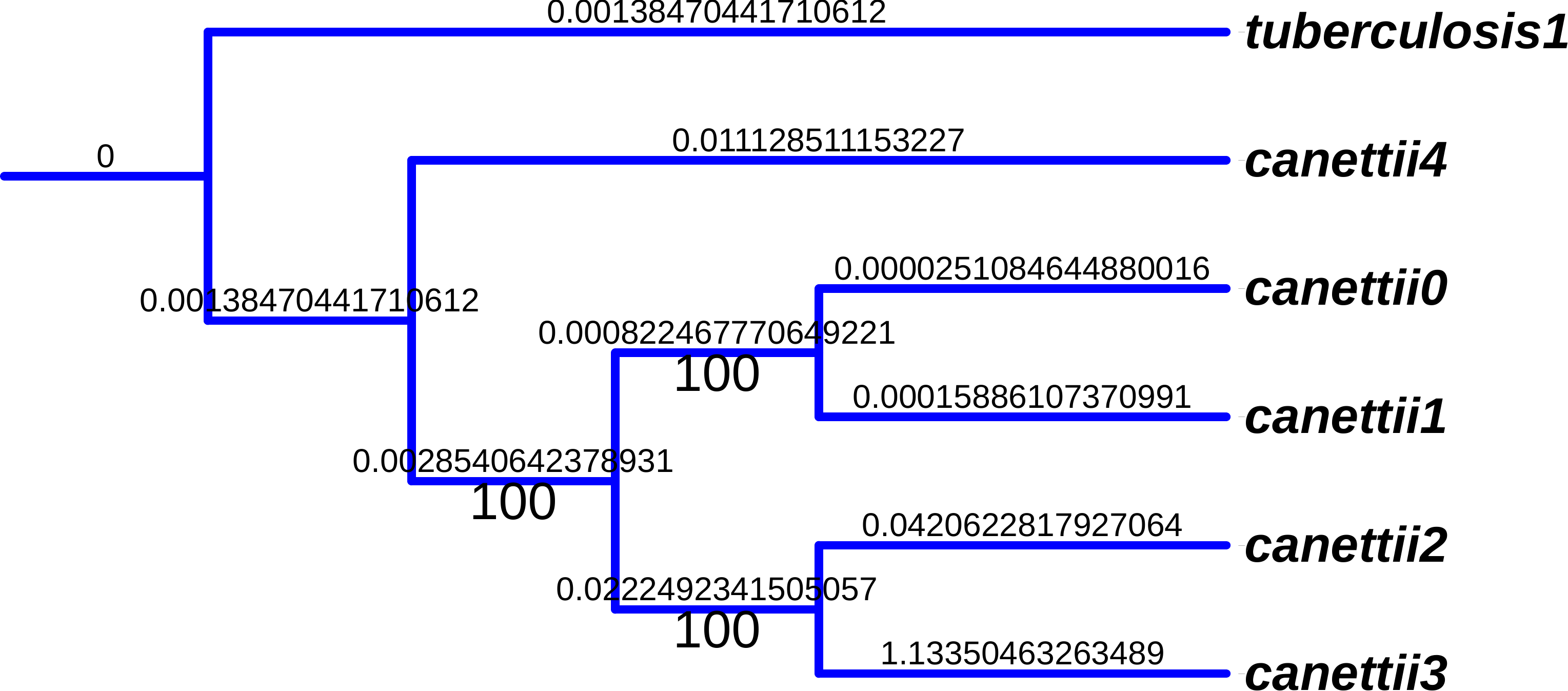}}
\end{minipage}%
\begin{minipage}{.5\linewidth}
\centering
\subfloat[]{\label{fig:M.tuberculosis_tree}\includegraphics[width=0.8\textwidth]{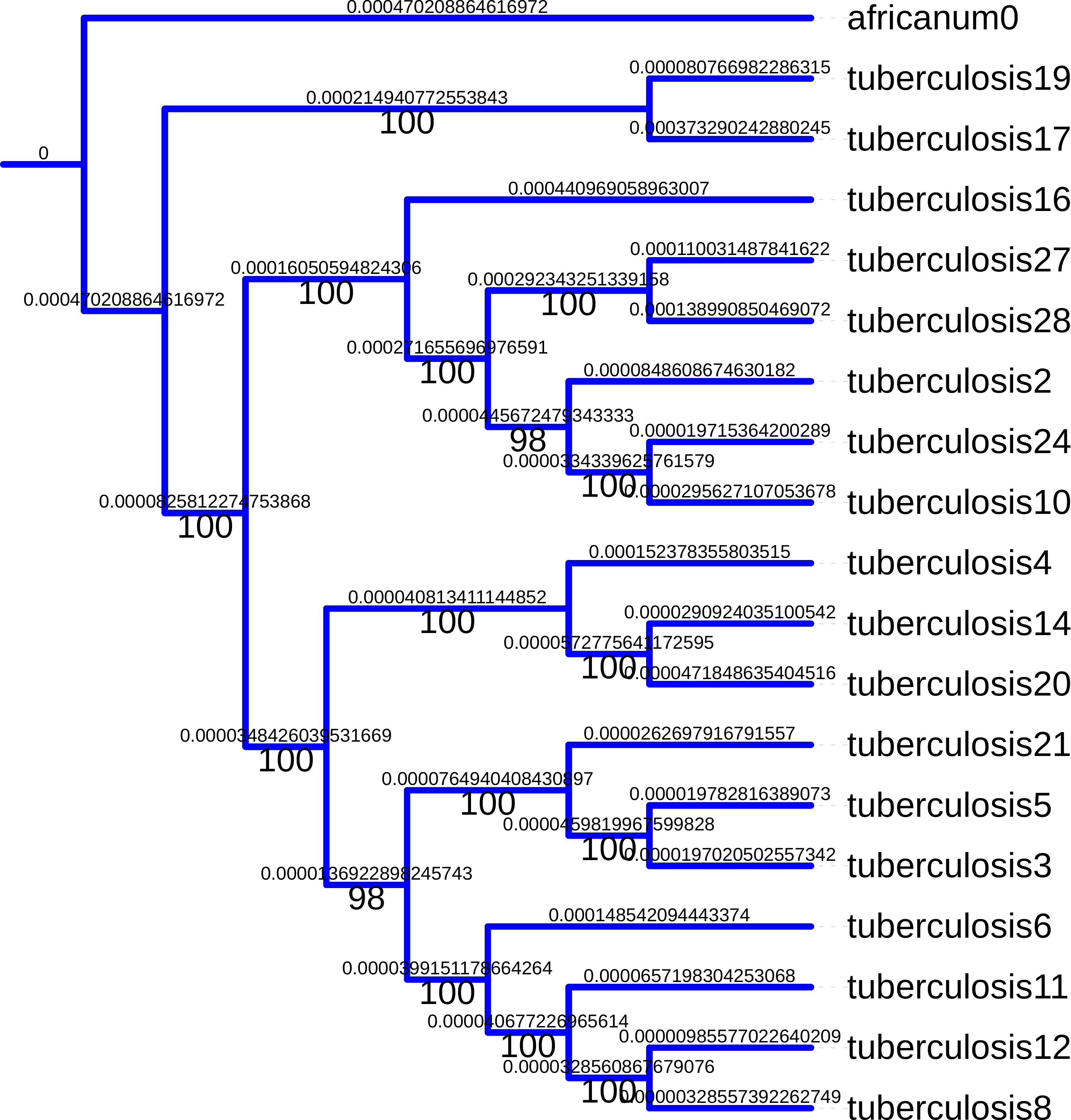}}
\end{minipage}\par\medskip
\centering
\caption{Well-supported phylogenies: (a) \textit{M. canettii} species using a \textit{M. tuberculosis} as outgroup, (b) \textit{M. tuberculosis} species with \textit{M. africanum} as outgroup. Phylogenetic trees have been calculated on the entire genomes with RAxML and GTR Gamma model.}
\end{figure}

As we reversed the inversions, our phylogenetic investigations are based on the whole genome.
This leads to well supported and trustworthy trees of strains, on which we can reliably consider to reconstruct ancestral states. As an illustrative example, we represent the phylogenetic trees of \textit{M. canettii} species with a relevant outgroup in Figure~\ref{fig:M.canettii_tree}. 
This very well supported tree has been obtained using RAxML  with GTR Gamma model as advised by JModelTest 2.0.
The \textit{Mycobacterium tuberculosis} phylogeny, for its part, leads to bootstrap supports larger than 98\%, as shown in Figure~\ref{fig:M.tuberculosis_tree}.

Having a confidential representation of the general evolution of MTBC strains due to this phylogenetic study, we are then left to reconstruct the ancestral states of the alignment at each internal node of the tree. This final ancestral reconstruction will be applied in two stages, considering first the variants of length 1 in the alignment (namely, single nucleotide polymorphism and indels of 1 nucleotide), and then larger variants that mainly consist of insertion or deletion of a subsequence at a location in the tree.

\subsection{Ancestral reconstruction: mononucleotidic variants}

Focusing on mononucleotidic variants, we separated the treatment of single nucleotide polymorphisms (SNPs) versus insertion-deletions (indels). For the former, the situation seems quite simple, the only problem being to prevent confusion between a ``true'' SNP and a SNP induced by a recombination of the indel kind. For the latter, future challenges encompass to determine which indels are related to tandem repeats, which are associated with mobile elements, or which are due to repeated sequences. Let us detail each case hereafter.

Regarding SNPs, the ancestral reconstruction is achieved as follows. The marginal probability distributions for bases at ancestral nodes in the phylogenetic tree are first calculated. These distributions are obtained using the sum-product message passing algorithm~\cite{pearl1982reverend}, assuming independence of sites. The ancestral reconstruction is done by using PHAST software~\cite{hubisz2010phast}, which reconstructs indels too by parsimony, also assuming site independence. Obtained results on mononucleotidic variants are then carefully visually checked, as the number of such variants is not excessive, see Tables~\ref{polymorphism_canettii3} 
and~\ref{polymorphism_tuberculosis}. 

\begin{figure}[!htb]
\begin{minipage}{.5\linewidth}
\centering
\subfloat[\textit{M. canettii} species]{ \label{fig:M.canettii_SNPsLocation}\includegraphics[width=0.8\textwidth]{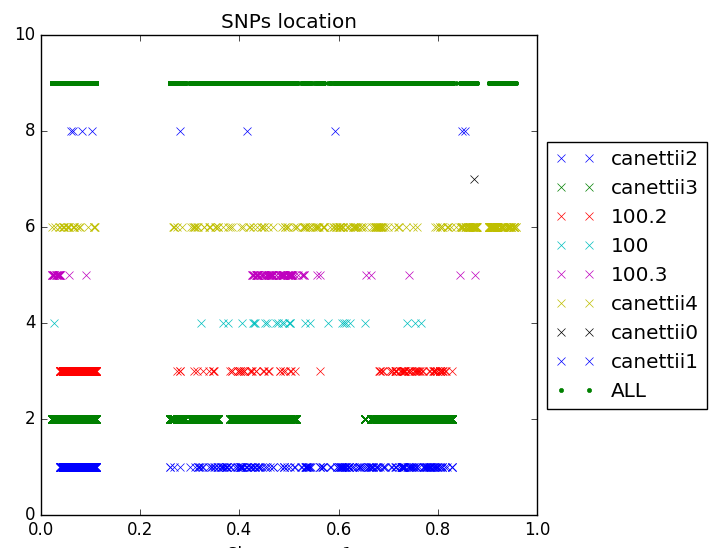}}
\end{minipage}%
\begin{minipage}{.5\linewidth}
\centering
\subfloat[\textit{M. tuberculosis} species]{ \label{fig:M.tuberculosis_SNPsLocation}\includegraphics[width=0.9\textwidth]{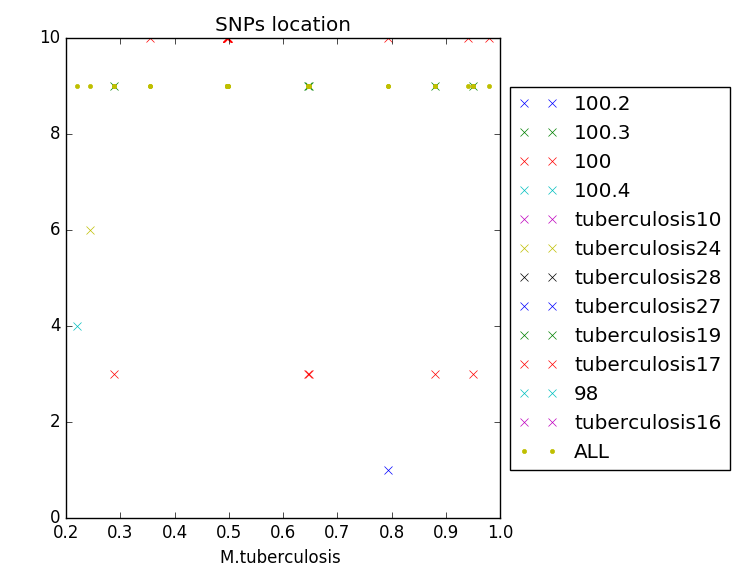}}
\end{minipage}\par\medskip
\centering
\caption{SNPs location of mononucleotidic variants}
\label{fig:SNPsLocation}
\end{figure}

At the end, 2,956 SNPs and 166 indels have been found in the alignment of the clade constituted by the 5 strains of \textit{M. canettii}, as shown in Figure \ref{fig:M.canettii_SNPsLocation}. Figure \ref{fig:M.tuberculosis_SNPsLocation}, for its part, represents the location of the 394 SNPs and of the 25 indels that have been found in the alignment of the clade constituted by 8 genomes of \textit{M. tuberculosis}.

\subsection{Ancestral reconstruction of larger variants}

\begin{figure}[!htb]
\begin{minipage}{.5\linewidth}
\centering
\subfloat[]{\label{fig:canettii_5}\includegraphics[width=0.8\textwidth]{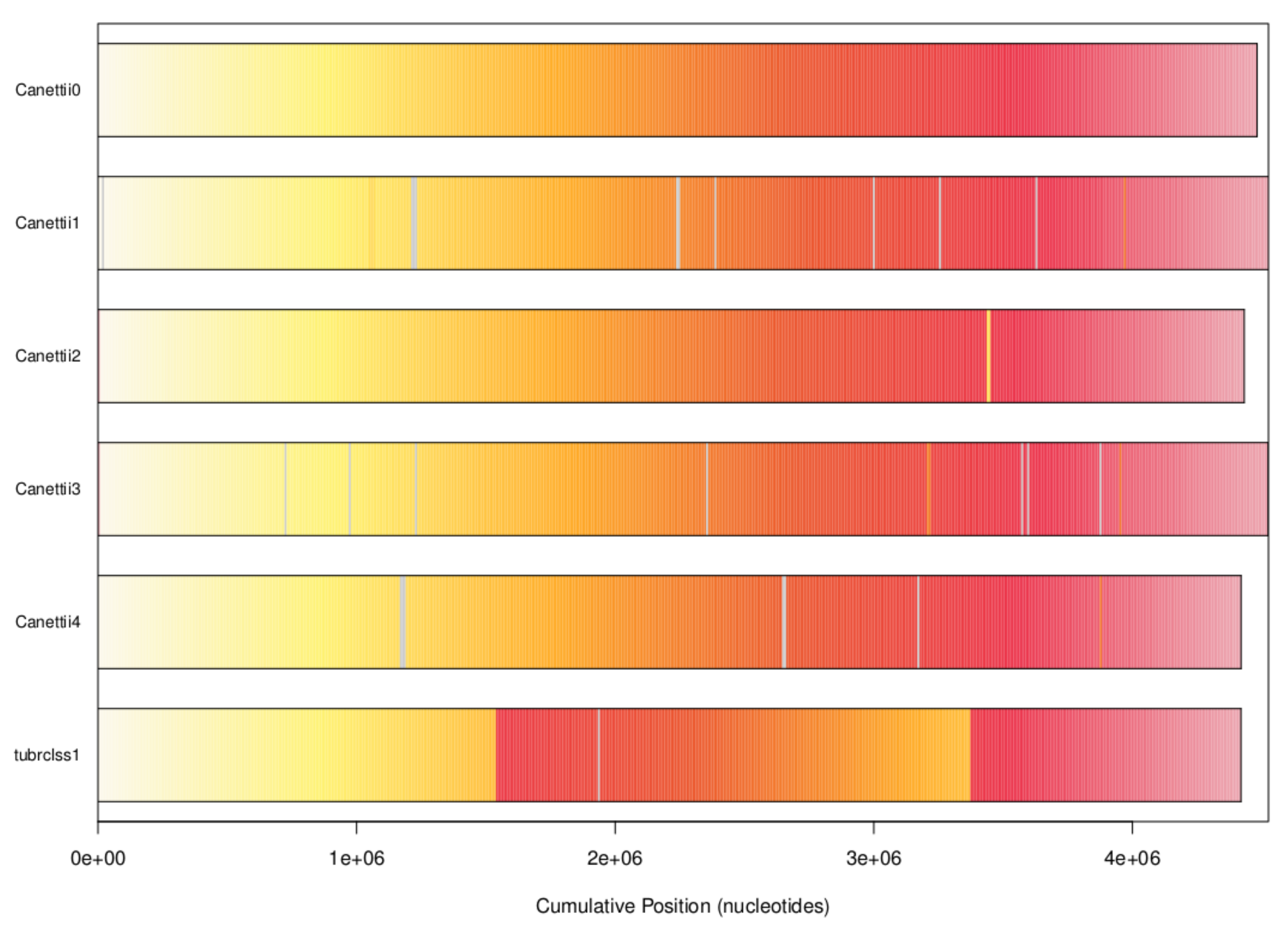}}
\end{minipage}%
\begin{minipage}{.5\linewidth}
\centering
\subfloat[]{\label{fig:canettii_6}\includegraphics[width=0.8\textwidth]{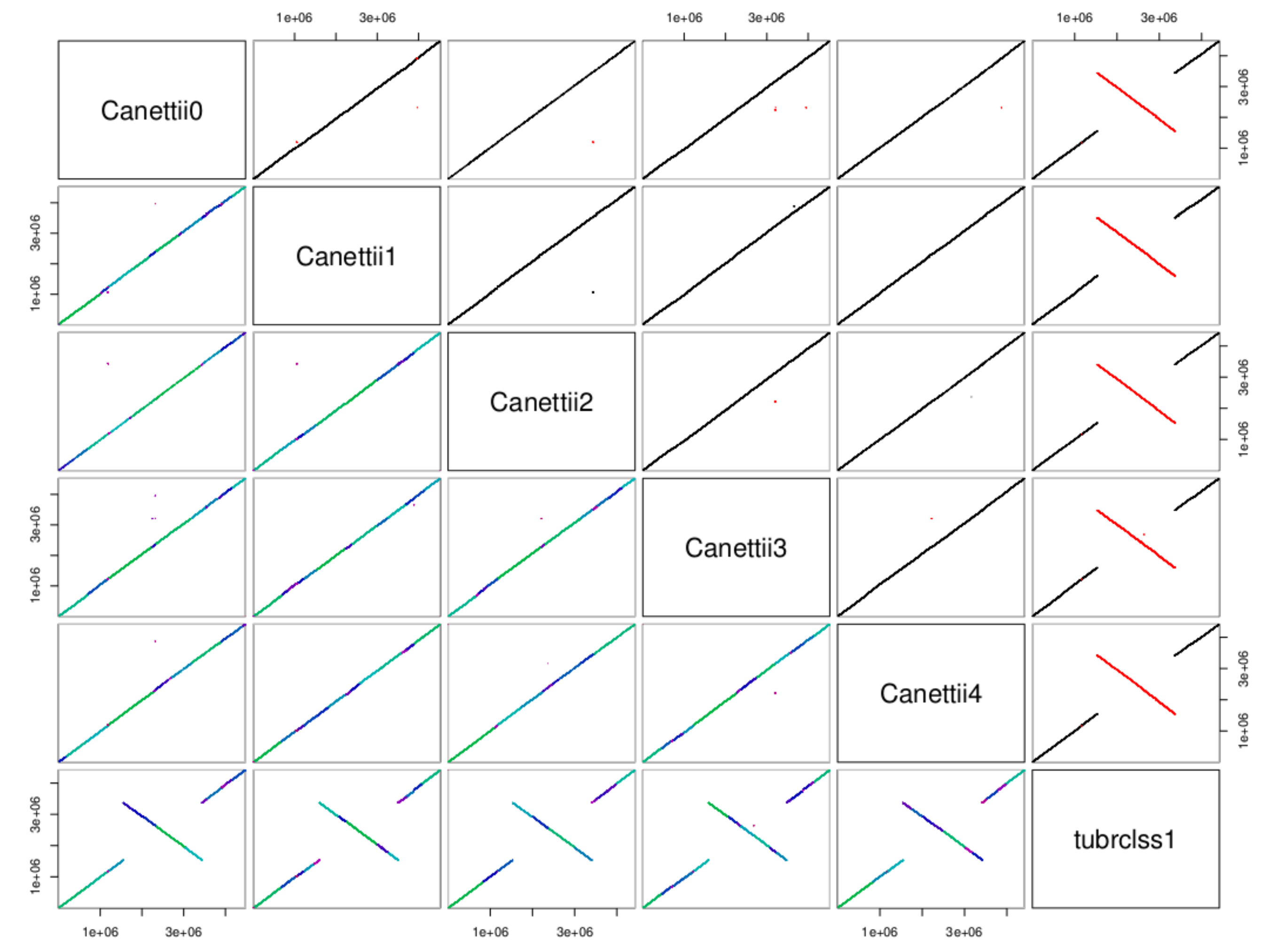}}
\end{minipage}\par\medskip
\centering
\caption{(a) synteny blocks in \textit{canettii}. Each genome is colored according to the position of the corresponding region in the first genome (gray if a region is unshared).
(b) Dot plots provide an alternative representation of the synteny map of~\textit{M.  canettii}. Black diagonal lines show syntenic regions sharing the same orientation, whereas red anti-diagonal ones represent blocks of synteny between opposite strands.
The description of all of these species tends to show a high sequence similarity with little recombination events.}
\label{fig:canettii_Barplot}
\end{figure}

\begin{figure}[!htb]
\centering
\includegraphics[width=13 cm]{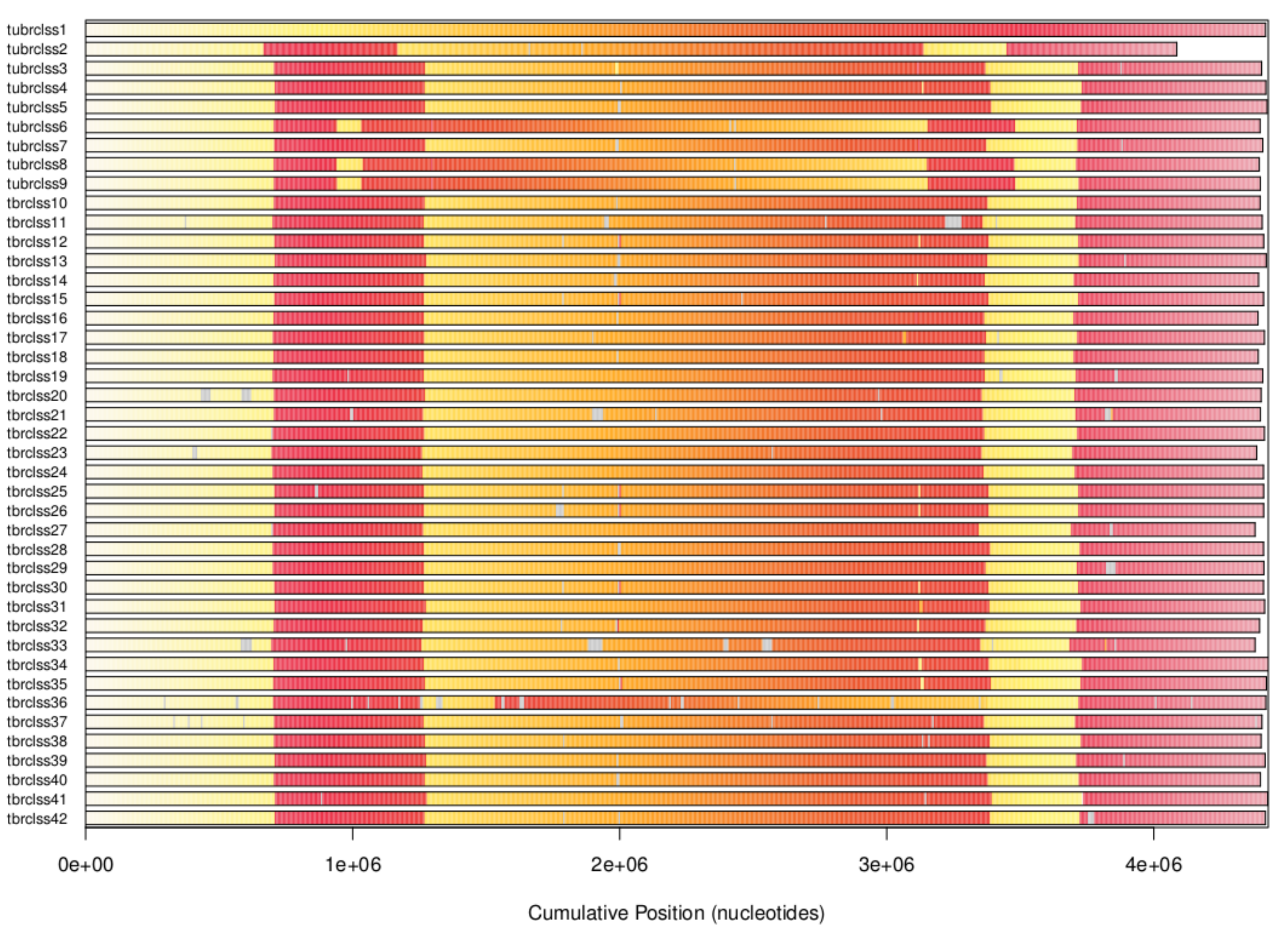}
\caption{A representation of ~\textit{M. tuberculosis} genomes species tends to show  more than $95\%$ nucleotide similarity with little recombination events.} 
\label{fig:M.tuberculosis_1}
\end{figure}

\textit{Mycobacterium} species considered in this article are highly conserved, with really similar regions and without rearrangement. 
As previously evoked, we found only a few significant inversions, like the one at the last common ancestor of strains \textit{CIPT 140010059}, \textit{140070010}, 
\textit{140060008}, \textit{140070017}, and \textit{140070008}, as shown in Figure~\ref{fig:canettii_5}.
Figure~\ref{fig:canettii_6}, for its part, is a dotplot representing these homologous regions, as identified by the FindSynteny function in R. synteny blocks of the 42 \textit{M. tuberculosis} are finally depicted in Figure~\ref{fig:M.tuberculosis_1}, where we have obtained 99\% of DNA sequence identity.
To sum up, if we except a large scale inversion, we can only report some small indels at this recombination level.

Ad hoc algorithms have then been designed to deal with mid size variants. More specifically, we have written first a string algorithm that detect small and noisy inversions, but the latter, distributed on our supercomputer facilities, was only able to detect artifacts. So either the MTBC genomes have not faced inversion events during its recent history, or this recombination case still needs further investigations. Authors tend to prefer the first possibility, as \textit{Mycobacterium} genomes evolve in a clonal manner (which is not the case, for instance, with \textit{Yersinia} genus, in which a large amount of mobile elements has led to a large number of reported inversions) \cite{behr2013evolution}. Duplication, for its part, has not yet been investigated but, as for inversions, the analysis of synteny blocks tends to show that such events are rare, at least if we consider the large scale ones.

\begin{figure}[!htb]
\begin{minipage}{.5\linewidth}
\centering
\subfloat[]{\label{fig:canettii_gaps}\includegraphics[width=0.7\textwidth]{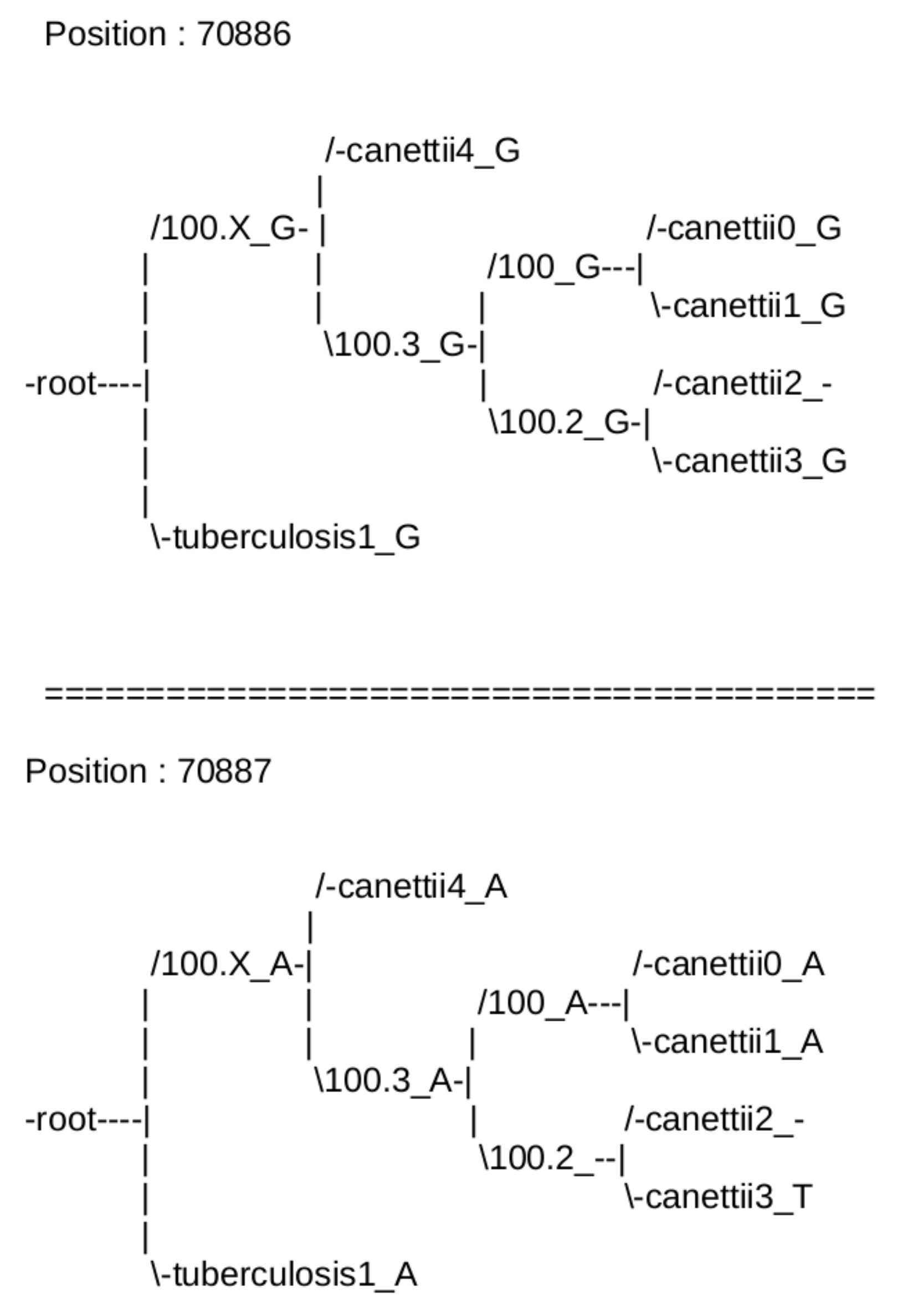}}
\end{minipage}%
\begin{minipage}{.5\linewidth}
\centering
\subfloat[]{\label{fig:tuberculosis_gaps}\includegraphics[width=0.7\textwidth]{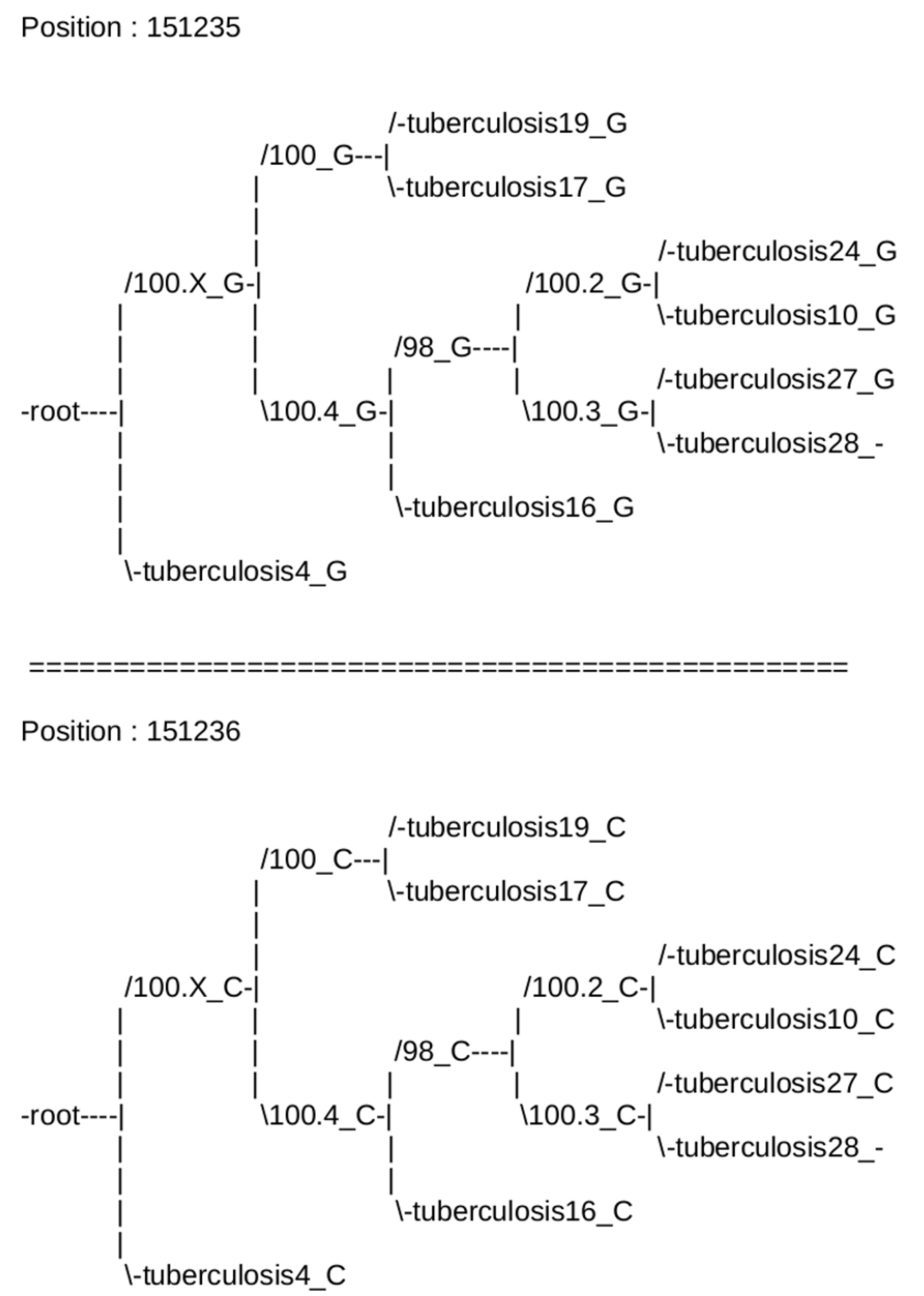}}
\end{minipage}\par\medskip
\centering
\caption{The insertions and deletions of nucleotides (indels) on the internal node of the tree (a)~represent the nucleotides contain the ancestor nodes and their children on \textit{M. canettii} species, (b)~\textit{M. tuberculosis} species.}
\label{fig:lesIndels1}
\end{figure}

Both indels of midsize and SNPs are rare, for its part, has been deeply studied, using PHAST software as detection tool. From obtained results, we can conclude the following points. (1) Such events are quite rare in some lineages of the MTB complex like \textit{tuberculosis}, as described in Table~\ref{table:snps}. (2) Most of the times, the situation is very easy to understand manually, leading either to an insertion or to a deletion at an obvious internal node of the tree, as illustrated in Figure~\ref{fig:lesIndels1}. (3) Most of the times, the inserted motif has not faced mutations during evolution: leaves that contain the motif have no mutation in it, thereby contributing to an easy to resolve situation. (4) Surprisingly,  ancestral states recovered by PHAST and its parsimony approach leads to disappointing results. Similarly, obviously wrong results have been obtained with state-of-the-art competitor software. To sum up, a manual reconstruction of mid size indels is possible, due to the low number of these recombinations that are mainly very easy to resolve, while automatic tools from the literature are not currently able to do it.


\begin{table}[!htb]
\tiny
\centering
\caption{Number of alignment columns with polymorphism, by pair of strains, on \textit{M. canettii} genomes. Note that, when a large string is deleted at some location in the tree, all the characters of this deletion are counted here.}
\label{polymorphism_canettii3}
\begin{tabular}{|c|c|c|c|c|c|c|}
\hline
\textit{\textbf{}} & \textit{\textbf{canettii0}} & \textit{\textbf{canettii1}} & \textit{\textbf{canettii2}} & \textit{\textbf{canettii3}} & \textit{\textbf{canettii4}} & \textit{\textbf{tuberculosis1}} \\ \hline
\textit{\textbf{canettii0}} &0&3524&27256&60957&4833&3354\\ \hline
\textit{\textbf{canettii1}} &3524&0&27260&61233&7971&1150\\ \hline
\textit{\textbf{canettii2}} &27256&27260&0&62717&27468&27437 \\ \hline
\textit{\textbf{canettii3}} &60957&61233&62717&0&60987&61346  \\ \hline
\textit{\textbf{canettii4}} &4833&7971&27468&60987&0&7510   \\ \hline
\textit{\textbf{tuberculosis1}} &3354&1150&27437&61346&7510&0   \\ \hline
\end{tabular}
\end{table}

\begin{table}[!htb]
\tiny
\centering
\caption{Variations in the alignment of \textit{M. tuberculosis}}
\label{polymorphism_tuberculosis}
\begin{tabular}{|c|c|c|c|c|c|c|c|c|}
\hline
\textit{\textbf{}} & \textit{\textbf{tuberculosis4}} & \textit{\textbf{tuberculosis19}} & \textit{\textbf{tuberculosis17}} & \textit{\textbf{\textit{tuberculosis16}}} & \textit{\textbf{tuberculosis27}} & \textit{\textbf{tuberculosis28}} & \textit{\textbf{tuberculosis24}} & \textit{\textbf{tuberculosis10}}\\ \hline
\textit{\textbf{tuberculosis4}} &0&199770&214401&219205&216387&217235&216919&217186\\ \hline
\textit{\textbf{tuberculosis19}} &199770&0&212403&219039&216908&216672&216726&216953\\ \hline
\textit{\textbf{tuberculosis17}} &214401&212403&0&216808&216534&217011&216786&216882 \\ \hline
\textit{\textbf{tuberculosis16}}&219205&219039&216808&0&216669&216916&216251&216678  \\ \hline
\textit{\textbf{tuberculosis27}} &216387&216908&216534&216669&0&142974&189148&199505  \\ \hline
\textit{\textbf{tuberculosis28}} &217235&216672&217011&216916&142974&0&189460&199412   \\ \hline
\textit{\textbf{tuberculosis24}} &216919&216726&216786&216251&189148&189460&0&194315   \\ \hline
\textit{\textbf{tuberculosis10}} &217186&216953&216882&216678&199505&199412&194315&0   \\ \hline
\end{tabular}
\end{table}

\begin{figure}[!htb]
\begin{minipage}{.5\linewidth}
\centering
\subfloat[]{\label{fig:Ancestor_canettii}\includegraphics[width=0.7\textwidth]{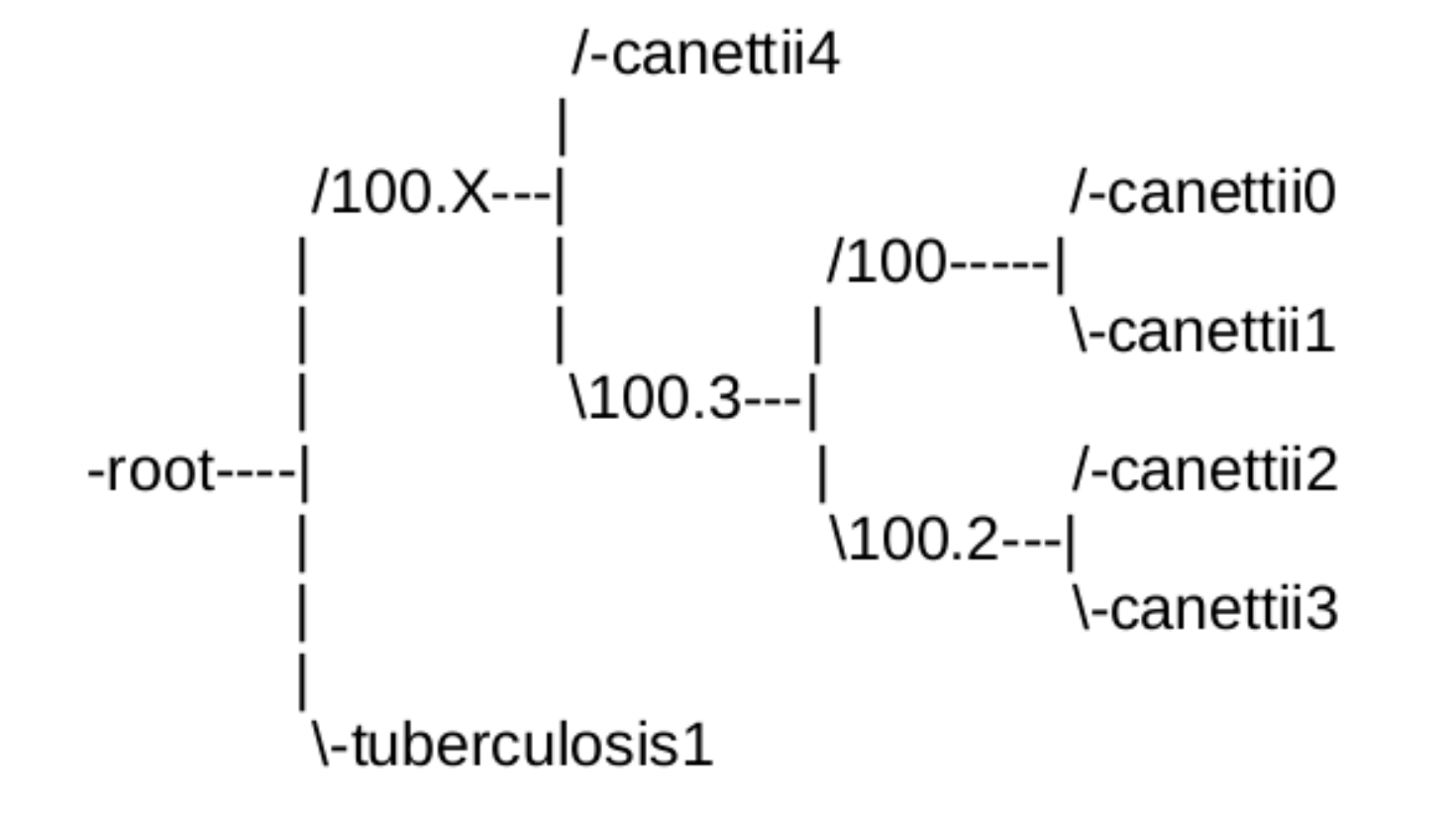}}
\end{minipage}%
\begin{minipage}{.5\linewidth}
\centering
\subfloat[]{\label{fig:Ancestor_tuberculosis}\includegraphics[width=0.7\textwidth]{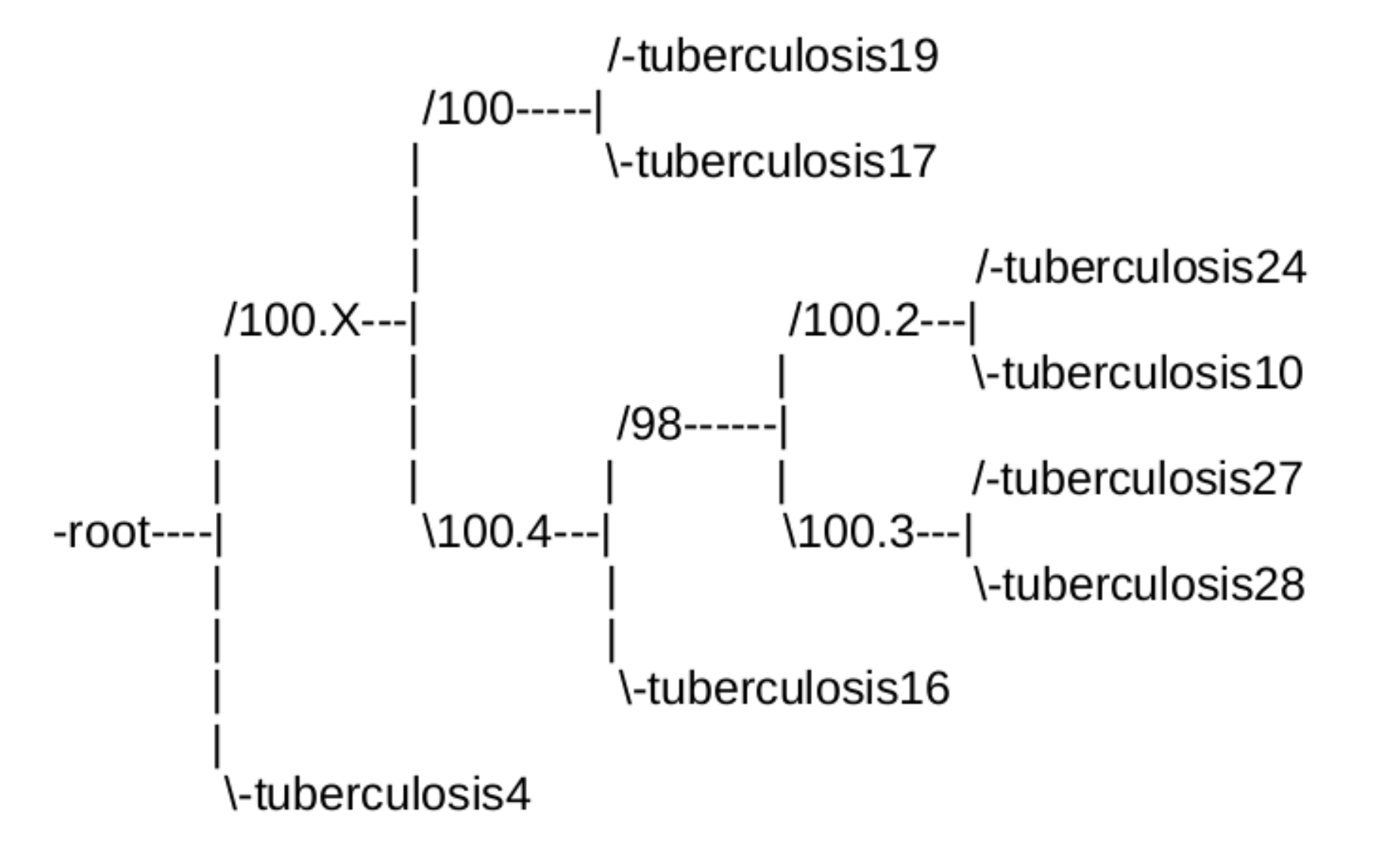}}
\end{minipage}\par\medskip
\centering
\caption{Example of phylogenetic tree, show the ancestor nodes (internal node), and the relation with their children (a) \textit{M. canettii} species, (b) \textit{M. tuberculosis} species.}
\label{fig:Ancestor_tree}
\end{figure}

\begin{table}[!htb]
    \begin{minipage}{.5\linewidth}
    \tiny
      \centering
      \textit{M. canettii} SNPs\\
        \begin{tabular}{|c|c|c|}
        \hline
        \textbf{Fathers}                & \textbf{Children}  & \textbf{No. of SNPs} \\ \hline
        \multirow{2}{*}{\textit{100.2}} & \textit{canettii2} & 1041                   \\ 
                                & \textit{canettii3}  & 12398                  \\ \hline
        \multirow{2}{*}{\textit{100}}   & \textit{canettii0}  & 1                   \\ 
                                & \textit{canettii1}  & 9                   \\ \hline
        \multirow{2}{*}{\textit{100.3}} & \textit{100}       & 28                   \\ 
                                & \textit{100.2}     & 735                   \\ \hline
        \multirow{2}{*}{\textit{100.X}} & \textit{100.3}     & 111                   \\ 
                                & canettii4          & 438                  \\ \hline
      \end{tabular}
    \end{minipage}%
    \begin{minipage}{.5\linewidth}
      \centering
      \tiny
        \textit{M. tuberculosis} SNPs\\        \begin{tabular}{|c|c|c|}
          \hline
           \textbf{Fathers}                & \textbf{Children}  & \textbf{No. of SNPs} \\ \hline
           \multirow{2}{*}{\textit{100}} & \textit{tuberculosis19} & 5                   \\ 
                                & \textit{tuberculosis17}  & 14                 \\ \hline
           
            \multirow{2}{*}{\textit{100.2}}   & \textit{tuberculosis24}  & 1                   \\ 
                                  & \textit{tuberculosis10}  & 0                   \\ \hline 
             \multirow{2}{*}{\textit{100.3}} & \textit{tuberculosis27}       & 0                   \\ 
                                & \textit{tuberculosis28}     & 0                 \\ \hline                     
             \multirow{2}{*}{\textit{98}}   & \textit{100.2}  & 1                   \\ 
                                & \textit{100.3}  & 0                   \\ \hline                     
             \multirow{2}{*}{\textit{100.4}}   & \textit{98}  & 0                   \\ 
                                  & \textit{tuberculosis16}  & 1                   \\ \hline

               \multirow{2}{*}{\textit{100.X}} & 100     & 5                   \\ 
                                & 100.4   & 1                 \\ \hline
         \end{tabular}
    \end{minipage} 
    \caption{Number of SNPs in the considered species (100.X refers to an ancestral node, as in the tree)}
    \label{table:snps}
\end{table}

\begin{figure}[!htb]
\centering
\includegraphics[width=0.5\textwidth]{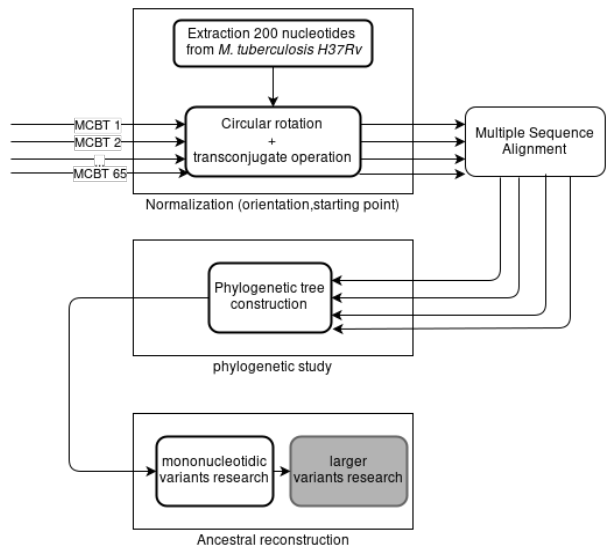}
\caption{Flowchart of the proposed approach.}
\label{fig:flowchart}
\end{figure}

All these steps are summarized in figure~\ref{fig:flowchart}. In this one, gray boxes correspond to manual steps whereas all the other ones are automatically executed.

\section{Discussion}
\label{results}
The obtained ancestors have not yet been studied in this work-in-progress. They will be investigated with updated and improved algorithms, encompassing mobile elements and gene content evolution analyzes.

Indeed, an important category of genome modification is the loss of functional genes, for instance because they become ineffective. In order to do so, we will consider the phylogenetic tree whose leaves will contain sets of genes, and we will compute core and pan genomes at each internal node of the tree. Having this core and pan tree, we will design an algorithm to investigate more deeply the evolution of these pan and core genomes over the tree, to see if some branches can be related to hot spots of evolution. We thus intend to determine at which rate such loss or gain occur, and which kinds of functionality are concerned. 
We will finally compute how much mutations fall inside a coding sequence, by studying which kind of genes has evolved on the phylogenetic tree, by wondering if the mutation rate has changed over time, and if such mutability can be related to environmental events. In other words, we will wonder which variations have been potentially significant among the numerous events that have been found when aligning these sequences.

With such a pipeline, we intend to investigate the following questions. Are some recombinations at the origin of severe tuberculosis epidemics? Are transposases responsible of such recombinations like inversions~\cite{siguier2006insertion,bergman2007discovering}? Are transposases in general more present in \textit{M. tuberculosis} (affecting humans) than in \textit{M.  africanum}, \textit{M. bovis}, or \textit{M. bovis BCG}? Are they related to the virulence of the strain? How core and pan genomes have evolved over time in this complex? Finally, we will compare the last common ancestor of this complex to a \textit{M. canettii}, to see if the \textit{canettii} ancestor hypothesis can be verified by the ancestral reconstruction way.

At this point, our partial conclusion is that the reconstruction of ancestral sequences is possible, at least in the case of close and clonal bacterias. Furthermore, elements being part of this reconstruction have already be designed, at least in their first revision (for instance to detect and deal with mononucleotidic variants). However, the MTB complex seems to be a little too complicated for a first deep investigation of semi-automatic reconstruction of ancestral sequences of bacteria, and a genus like \textit{Brucella} may be more easy to deal with in a first concrete investigation of this problem.

\section{Conclusion}
In this article, we have firstly emphasized that, even if various algorithms and software already exist to face the NP-hard character of the ancestral genome reconstruction problem, they do not work perfectly, in particular when SNPs or indels fall into repeated sequences. We have then argued that, when regarding the relatively low number of mutation and recombination events in such \textit{Mycobacterium}, a pragmatic approach is possible. 
We have proposed to reconstruct all ancestors of all complete available genomes of \textit{Mycobacterium tuberculosis} and of  \textit{M. canettii}. The study has started by investigating single nucleotide polymorphism level, while indels and large scale recombination are regarded in a second stage. 
Our conclusion is that, by mixing automatic reconstruction of obvious situations with human interventions on signaled problematic cases, it may be possible to achieve a concrete, complete, and really accurate reconstruction of some specific bacteria lineages.
We can thus investigate how these genomes have evolved from their last common ancestors. 

In future work, we intend to reconstruct all ancestors of all complete available genomes of specific bacteria strains, namely, and ordered by complexity: \textit{Brucella} genus, \textit{Yersinia pestis}, and \textit{Pseudomonas aeruginosa}. Moreover, we intend to compare them with ancient DNA when available (like for \textit{Y. pestis}). In parallel, original mathematical description of some recombination mechanisms will be proposed, encompassing branching process and partial differential equation approaches for modeling mobile elements. Finally, we may try to correlate the evolutionary history of microorganisms to epidemiological data: events of genomic recombination may be related to epidemic outbreaks. And such putative correlations may be learnt by deep learning algorithms, leading to a new way to predict epidemic risks.

\section*{Acknowledgments}

Computations presented in this article were realised
on  the  supercomputing  facilities  provided  by
the M\'esocentre de calcul de Franche-Comt\'e.

\bibliographystyle{unsrt}
\bibliography{Ref}
\end{document}